\documentclass[11pt,twoside]{article}
\usepackage{CAGN2016}
\usepackage{graphicx}

\usepackage[T1]{fontenc} 

\usepackage{latexsym}
\usepackage{verbatim}

\begin{document}

\vskip 1.0cm
\markboth{C.Kehrig}{Spatially resolved extremely metal-poor, high-ionizing SF galaxies}
\pagestyle{myheadings}
%
%
\vspace*{0.5cm}
\parindent 0pt{Invited Review}


\vspace*{0.5cm}
\title{Spatially resolved properties for extremely metal-poor star-forming
  galaxies with Wolf-Rayet features and high-ionization lines}

\author{C.Kehrig$^1$}
\affil{$^1$Instituto de Astrof\'{\i}sica de Andaluc\'{\i}a, CSIC, Apartado de correos 3004, E-18080 Granada, Spain}

\begin{abstract}

Extremely metal-poor, high-ionizing starbursts in the local Universe
provide unique laboratories for exploring in detail the physics of
high-redshift systems. Also, their ongoing star-formation and haphazard
morphology make them outstanding proxies for primordial
galaxies. Using integral field spectroscopy, we spatially resolved the
ISM properties and massive stars of two first-class low metallicity
galaxies with Wolf-Rayet features and nebular HeII emission: Mrk178
and IZw18. In this review, we summarize our main results for
these two objects.

\end{abstract}

\section{Introduction}

Local extremely metal-poor [i.e., 12+log (O/H) $\leq$
  7.7]\footnote{The precise value of the metallicity defining
  extremely metal-poor galaxies varies in the literature (see Guseva
  et al. 2016 and references therein)} starburt galaxies are among the
least chemically evolved objects in the nearby Universe, and are
considered to be analogues to the first star-forming (SF) systems
(e.g. Izotov et al. 1994, 2009; Hunter \& Hoffman 1999; Kehrig et
al. 2006; Sanchez Almeida et al. 2016).  Studying these
metal-deficient starbursts is needed to learn more about the evolution
and feedback from massive stars [e.g. Wolf-Rayet (WR) stars] in high-z
galaxies, and for exploring in detail the physics of the farway
Universe.

\begin{figure*} 
\begin{center}
\includegraphics[width=5.5cm,clip]{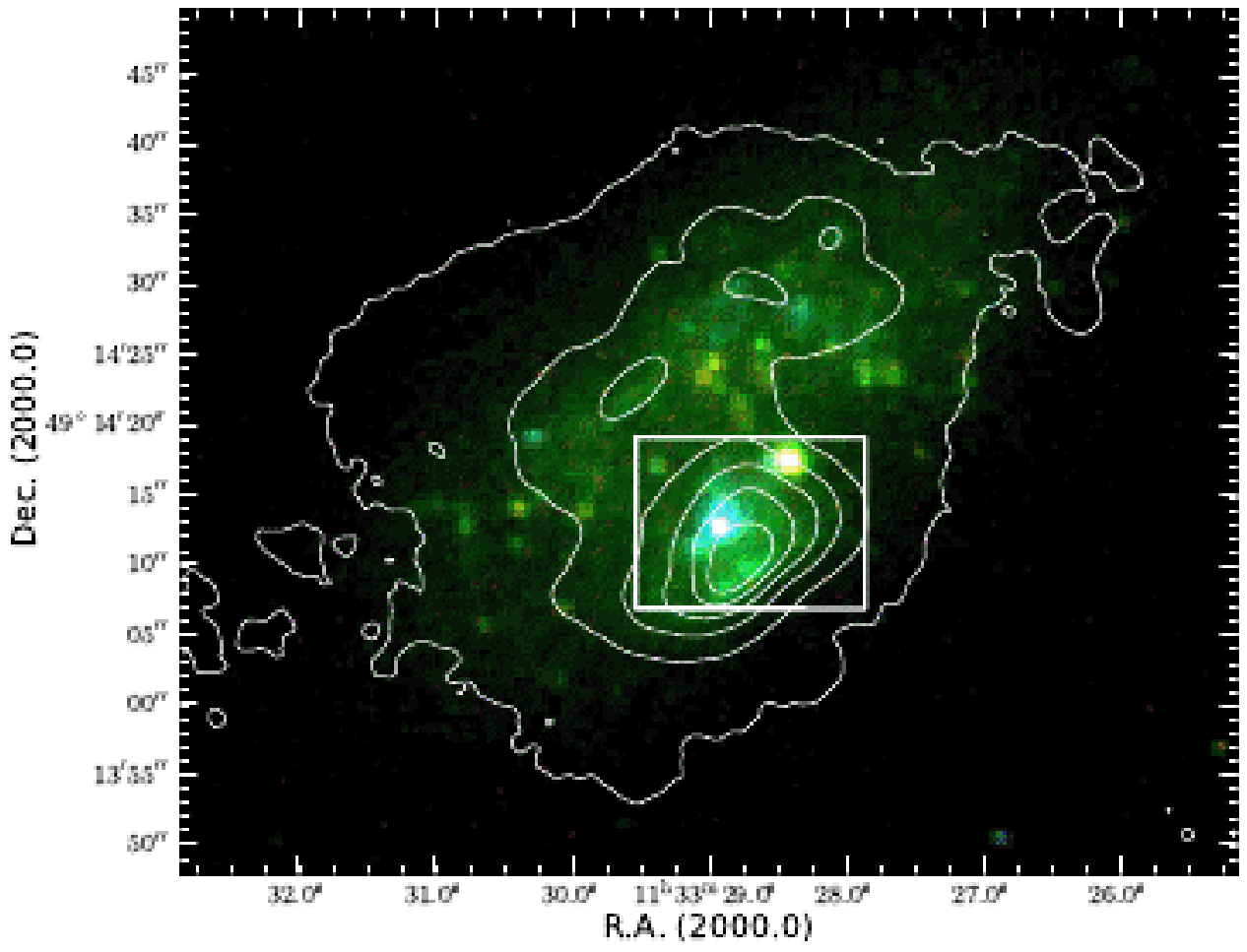}
\includegraphics[width=4.5cm,clip]{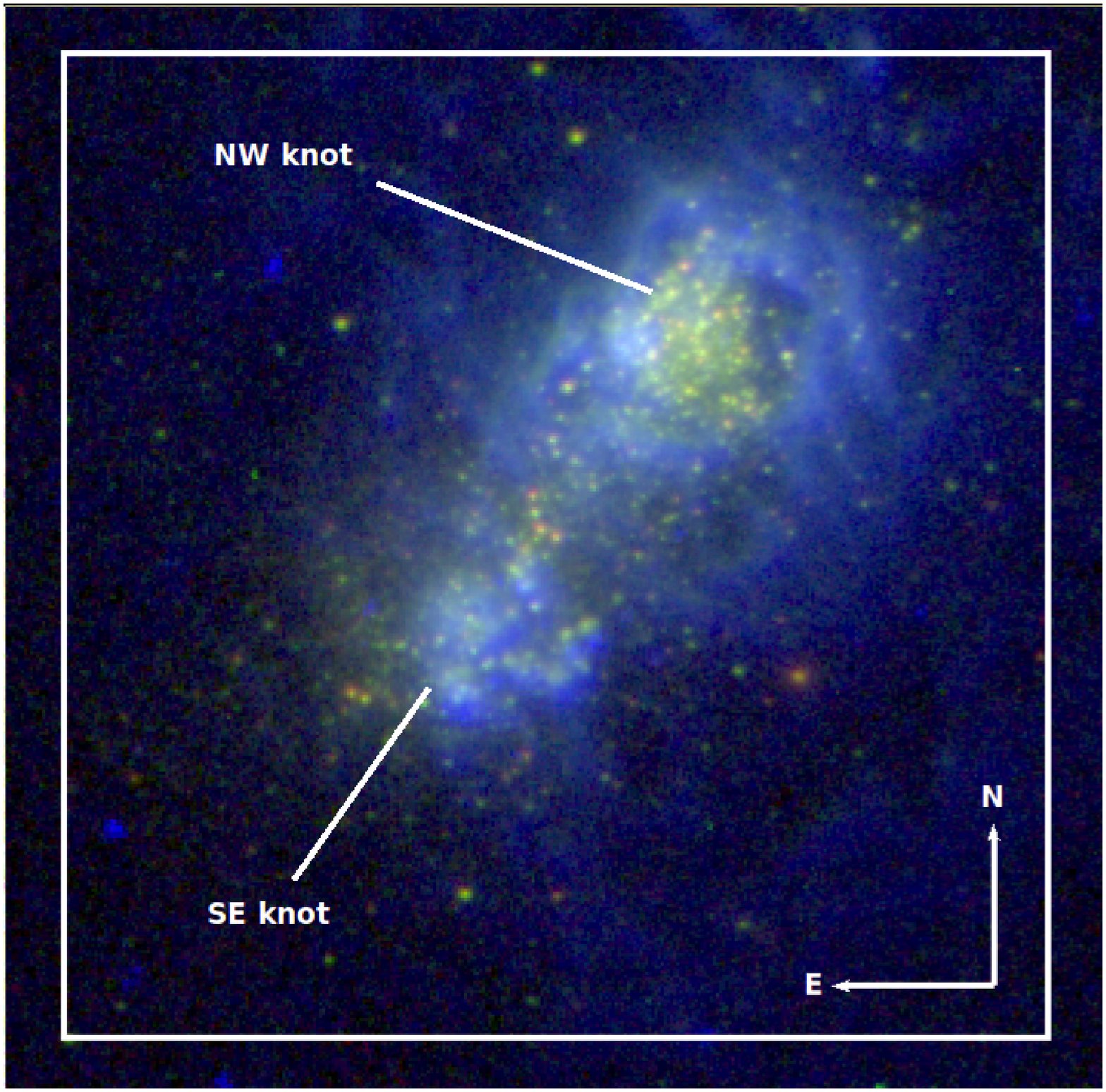}
\caption{{\em Left panel}: Colour-composite SDSS image of Mrk178
  overlaid with the observed FOV of WHT/INTEGRAL ($\sim$
  16''$\times$12'') represented by the white box. {\em Right panel}: Colour-composite HST image of IZw18. The observed FOV of PMAS (16''$\times$16'') is represented by the white box.}
\label{fov}
\end{center}
\end{figure*}

The presence of WR signatures (most commonly a broad feature centered
at $\sim$ 4680 \AA~ or ``blue bump'') in the spectra of some metal-poor SF
galaxies [e.g. Legrand et al. 1997; Guseva,
Izotov \& Thuan 2000; Cair\'os et al. 2010; Kehrig et al. 2016
(hereafter K16)] challenges current
single star (rotating/non-rotating) stellar evolution models that fail
in reproducing the WR content in low metallicity (Z) environments (see
Brinchmann, Kunth, \& Durret 2008 and references therein; Leitherer et
al. 2014). Thus, investigating the WR content and radiative feedback from WR stars
(WRs) in metal-poor starbursts is crucial to test the models at
low metallicity. The study on formation and thereabouts of gamma-ray bursts and
Type Ib/c SN progenitors, believed to be WRs in metal-poor galaxies,
may also benefit from the investigation presented here (e.g. Woosley
\& Bloom 2006)

The spectra of SF galaxies are dominated by strong nebular emission
lines which are mainly formed via the photoionization by hot massive
stars. High-ionization lines, like HeII, are often seen in the spectra
of low metallicity SF galaxies at both low and high redshift [e.g. Kehrig et
  al. 2004; Thuan \& Izotov 2005; Shirazi \& Brinchmann 2012; Cassata
  et al. 2013; Kehrig et al. 2013, 2015 (hereafter K13, K15)]. The
expected harder spectral energy distribution (SED) and higher nebular
gas temperatures at low metallicities should boost the supply of hard
ionizing photons. While the presence of hard radiation is well
established in some nearby metal-deficient SF galaxies, the
origin of this radiation is much less clear, in spite of several
attempts to account for it (e.g. Thuan \& Izotov 2005; K15).  Overall, several mechanisms for producing
hard ionizing radiation have been proposed, such as WR stars,
primordial zero-metallicity stars, high-mass X-ray binaries and fast
radiative shocks. However, no mechanism has emerged clearly as the
leading candidate. Reconsidering the underlying assumptions in the
analysis of high-ionization nebular emission in metal-poor galaxies is
key to advance our understanding of their properties.

We have used integral field spectroscopy (IFS) to obtain a more
believable view of extremely metal-poor, high-ionizing starbursts in
the local Universe (see Fig.\ref{fov}; K13, K15, K16). IFS has many advantages in
comparison with long-slit spectroscopy (e.g. Cair\'os et al. 2009;
James et al. 2011; Kehrig et al. 2012; P\'erez-Montero et al. 2013;
Duarte Puertas et al. 2016). IFS allows a more precise spatial correlation
between massive stars and nebular properties through a 2D analysis.
IFS is a powerful technique to probe and solve issues related with
aperture effects too.  Long-slit observations may fail in detecting
WR features due to their faintness with respect to the stellar
continuum emission and spatial distribution of WR stars across the
galaxy. In particular in low-Z objects, the dilution of WR features
and the difficulty in spectroscopically identifying WR stars is even
stronger owing to the steeper Z dependence of WR star winds which
lowers the line luminosities of such stars (e.g. Crowther \& Hadfield
2006). Kehrig et al. (2008) demonstrated for the first time the power of IFS in
minimizing the WR bump dilution and finding WR stars in extragalactic
systems where they were not detected before (see also Cair\'os et
al. 2010; James et al. 2011).

Here, we summarize the main results from the analysis of our new
integral field unit (IFU) data of two extremely metal-poor, high-ionizing SF
galaxies: Mrk178, {\it the closest metal-poor WR galaxy} (see K13), and IZw18,
{\it the most metal-deficient HeII-emitting SF galaxy known in the
  local Universe} (see K15, K16).

\section{Results}

\subsection{Mrk178: the closest metal-poor WR galaxy}

\begin{figure*}  
\begin{center}
\includegraphics[width=5.8cm,clip]{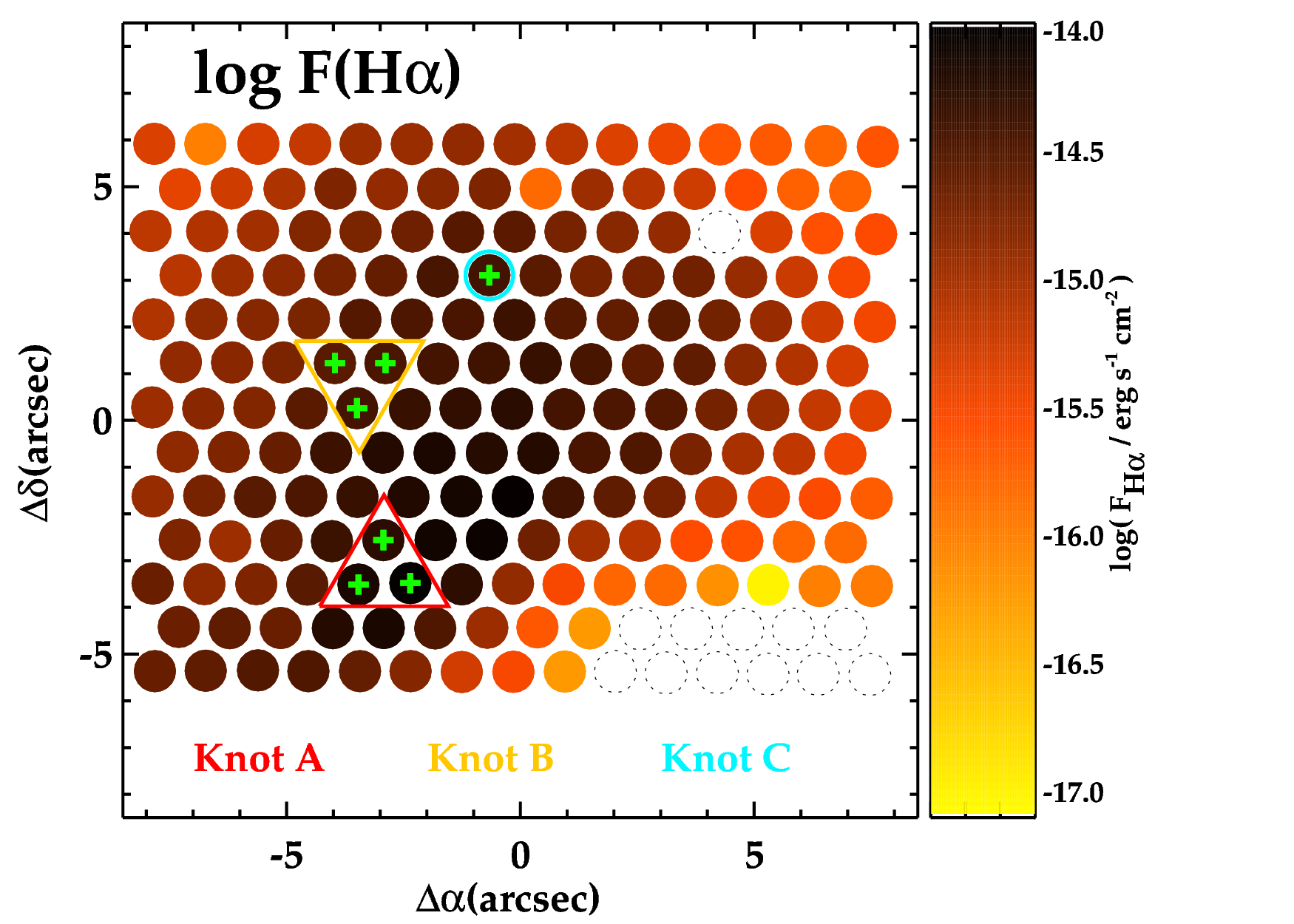}
\vspace{-0.25cm}
\includegraphics[width=5.5cm,clip]{kehrig_fig2b.eps}
\caption{{\it Left panel}: Map of H$\alpha$ emission line. The
  diameter of each spaxel is $\sim$ 1'' ($\sim$ 20 pc, our resolution
  element size). The three WR knots (A, B and C) are labelled on the H$\alpha$ flux map, and the spaxels where we detect WR features are marked with green crosses. {\it Right panel}: Integrated spectrum for the 3 knots in which WR features are detected. The spectral range for both blue and red WR bumps are marked (see K13).}
\label{mrk178.fig1}
\end{center}
\end{figure*}

\begin{figure*}
\begin{center}
\includegraphics[width=5.5cm,clip]{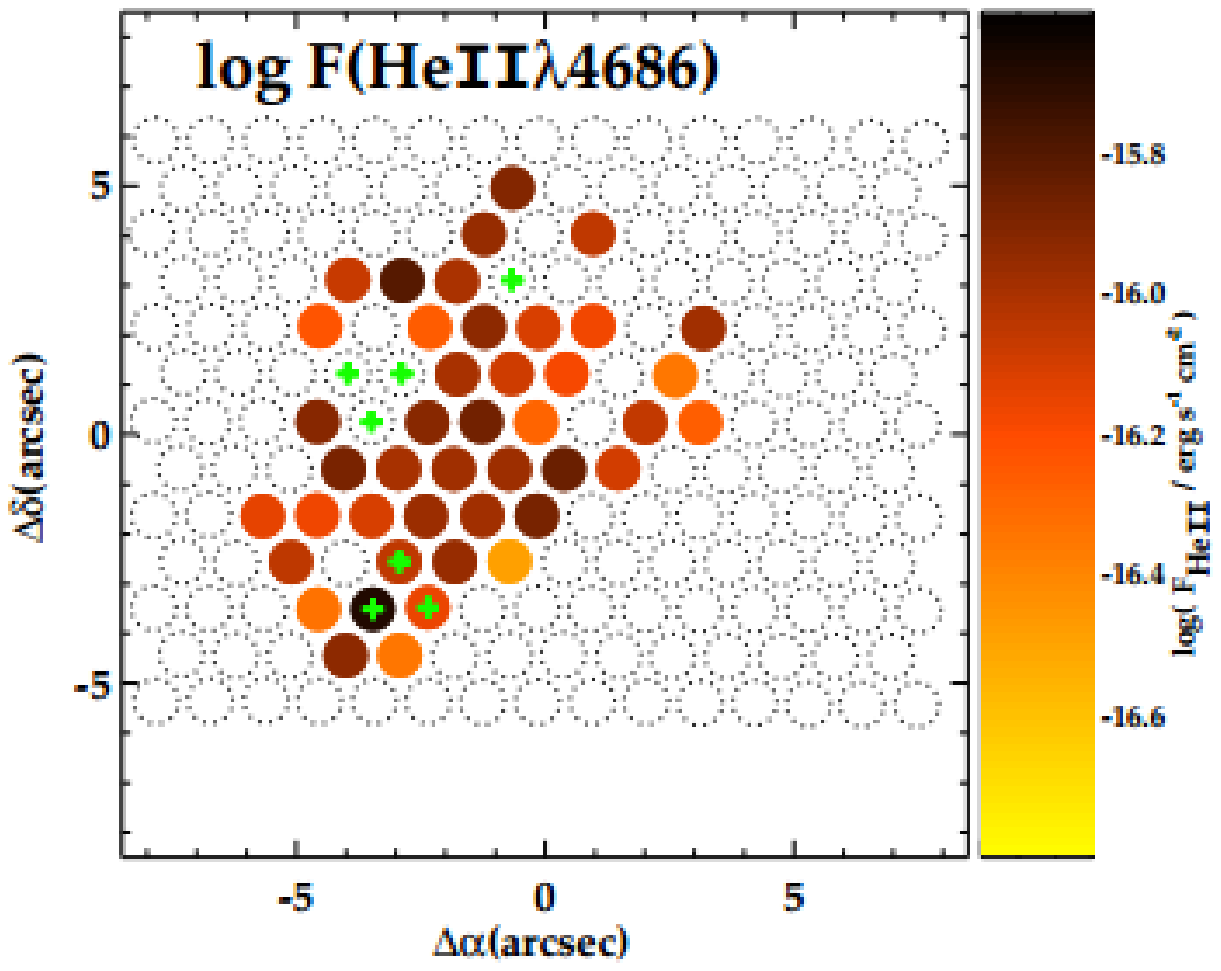}
\includegraphics[width=5.5cm,clip]{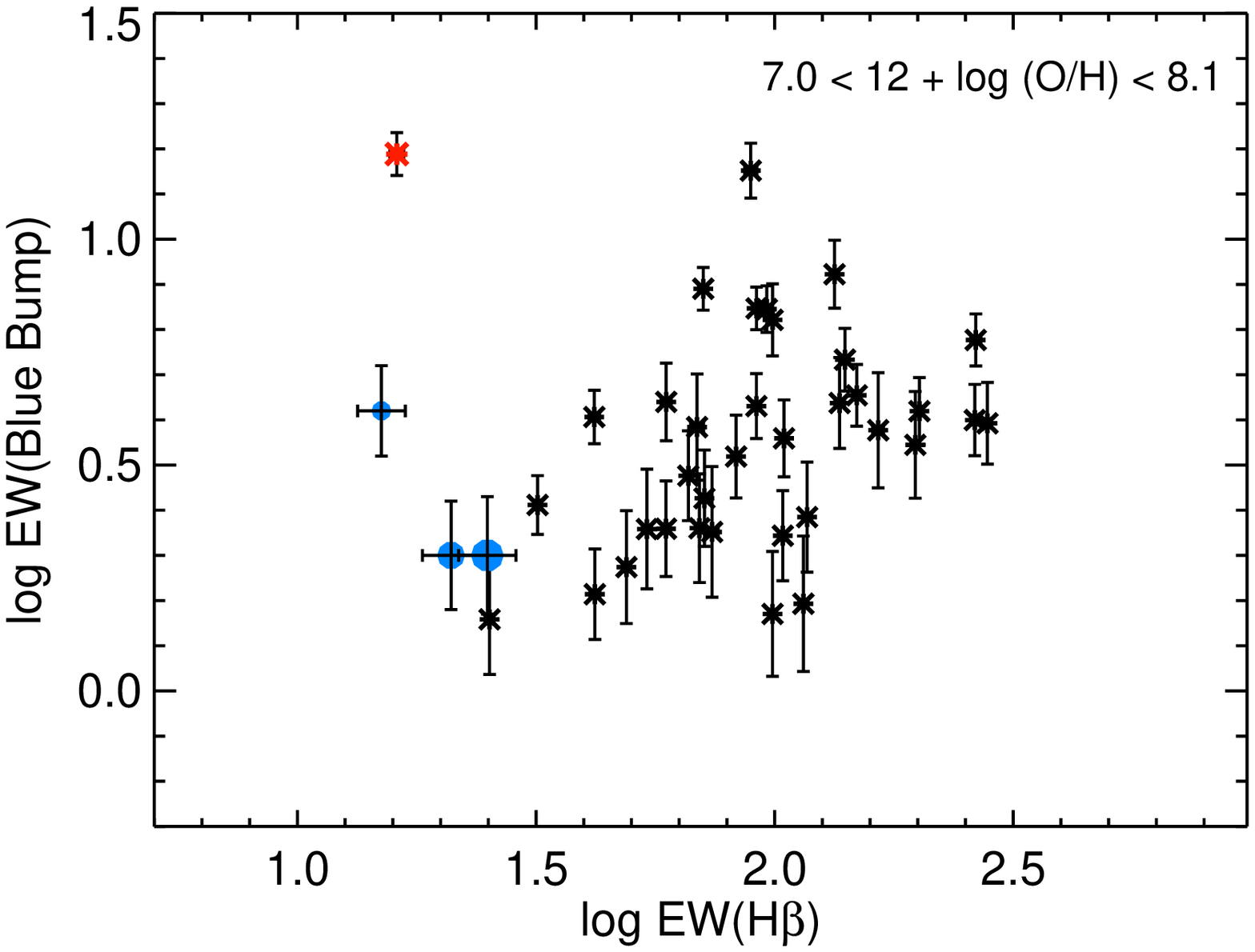} 
\caption{{\it Left panel}: map of nebular HeII$\lambda$4686 line; spaxels where
we detect WR features are marked with green crosses.
{\it Right panel}: EW(WR blue bump) vs. EW(H$\beta$). Asterisks show
values from SDSS DR7 for metal-poor WR galaxies; the red one
represents Mrk178. The three blue circles, from the smallest to the
biggest one, represent the 5'', 7'' and 10'' diameter apertures from
our IFU data centered at the SDSS fiber of Mrk178 (see K13 for details).
\label{mrk178.fig2}}
\end{center}
\end{figure*}

In K13 we present the first optical IFS study of Mrk~178 based on IFU
data obtained with the INTEGRAL IFU at the 4.2m WHT, Roque de los
Muchachos observatory (see Fig.\ref{fov}).  The proximity of Mrk~178
(distance $\sim$ 3.9 Mpc) combined with the IFS technique allow us to
locate and resolve SF knots hosting a few WRs, and also to
characterize the WR content.  In addition, we are able to probe the
spatial correlation between massive stars and the properties of the
surrounding ISM. 

We defined three WR knots from which two (knots A and C) are identified for the first
time in K13. The WR knot spectra reveal the presence of nitrogen-type
and carbon-type WR stars in Mrk~178 (see Fig.\ref{mrk178.fig1}). By
comparing the observed spectra of the WR knots with SMC/LMC template
WRs, we empirically estimate a lower limit for the number of WRs
($\geq$ 20) in our Mrk178 FOV that is already higher than that
currently found in the literature ($\sim$ 2-3 WRs from Guseva et
al. 2000). Regarding the ISM properties, our statistical analysis suggests that
spatial variations in the gaseous T$_{e}$[OIII] exist and that
the scatter in T$_{e}$[OIII] can be larger than that in O/H within the observed
FOV. Thus, caution should be exercised when analysing integrated
spectra of SF systems which do not necessarily represent the ``local''
ISM properties around massive star clusters. The nebular chemical
abundance in Mrk~178 is homogeneous over spatial scales of hundreds of
parsecs. The representative metallicity of Mrk~178 derived is
12+log(O/H) = 7.72 $\pm$ 0.01 (error-weighted mean value of O/H and
its corresponding statistical error). To probe the presence of
small-scale ($\sim$ 20 pc) localized chemical
variations, we performed a close inspection of the chemical abundances
for the WR knots from which we find a possible localized N and He
enrichment, spatially correlated with WR knot C (see also James et al. 2011). 

Nebular HeII$\lambda$4686 emission is shown to be spatially extended
reaching well beyond the location of the WRs (see HeII$\lambda$4686
map in Fig.\ref{mrk178.fig2}). Shock ionization and X-ray binaries are
unlikely to be significant ionizing mechanisms since Mrk178 is not
detected in X-rays, and measured values of [SII]/H$\alpha$ ($<$ 0.20)
are lower than the typical ones observed in SNRs. The main excitation source
of HeII in Mrk178 is still unknown.

From SDSS spectra of metal-poor WR galaxies, we found a too high EW(WR
bump)/EW(H$\beta$) value for Mrk178, which is the most deviant point
in the sample (see Fig.\ref{mrk178.fig2}).  Using our IFU data, we showed
that this curious behaviour is caused by aperture effects, which
actually affect, to some degree, the EW(WR bump) measurements for all
galaxies in left panel of Fig.\ref{mrk178.fig2}. Also, we demonstrated that using
too large an aperture, the chance of detecting WR features decreases.
This result indicates that WR galaxy samples constructed on
single fibre/long-slit spectrum basis may be biased in the sense that
WR signatures can escape detection depending on the distance of the
object and on the aperture size. 

\subsection{IZw18: the most metal-deficient HeII-emitting SF galaxy in the nearby Universe}

\begin{figure*}  
\begin{center}
\includegraphics[width=5cm,clip]{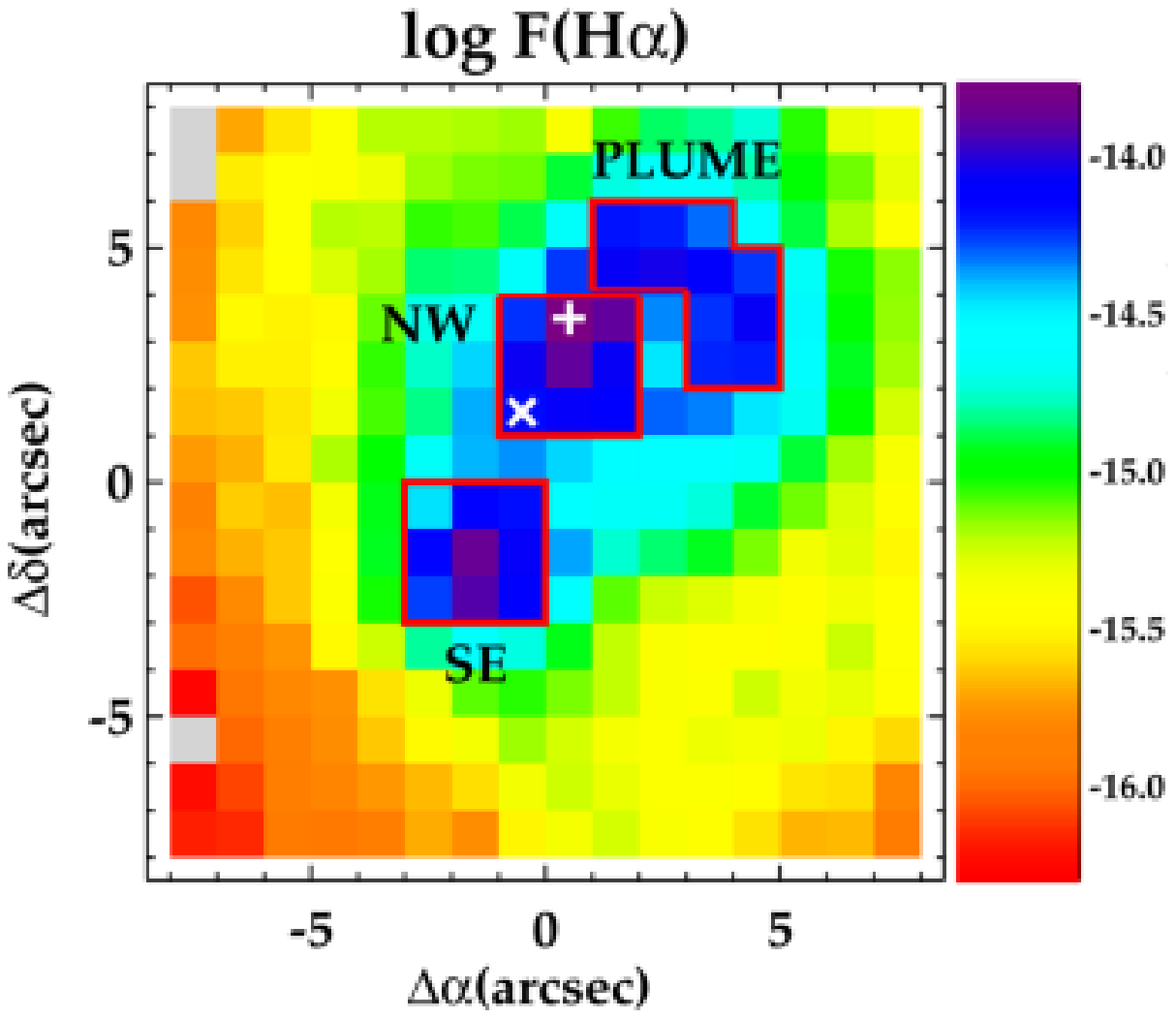}
\includegraphics[width=5cm,clip]{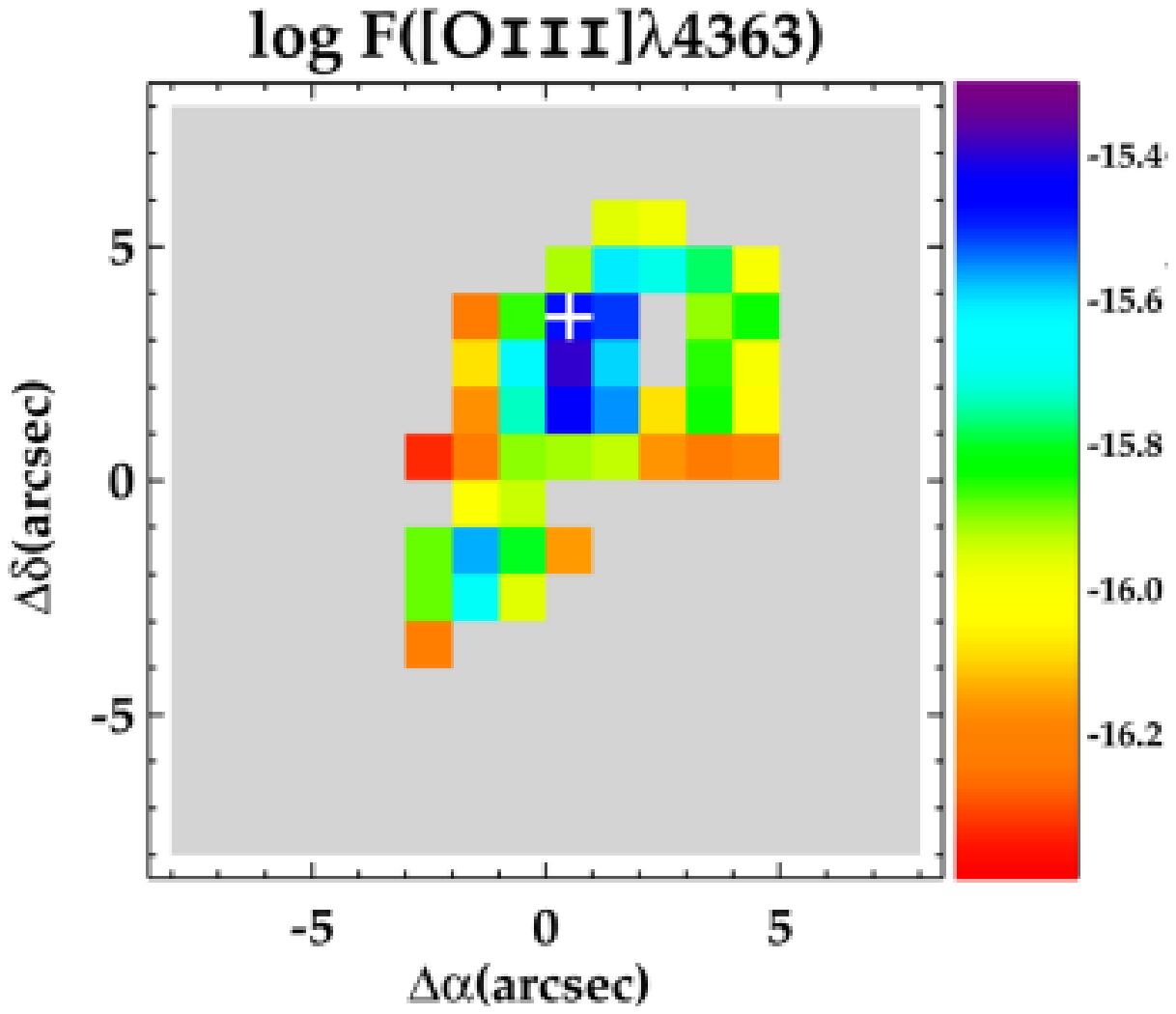}\\
\includegraphics[width=5cm,clip]{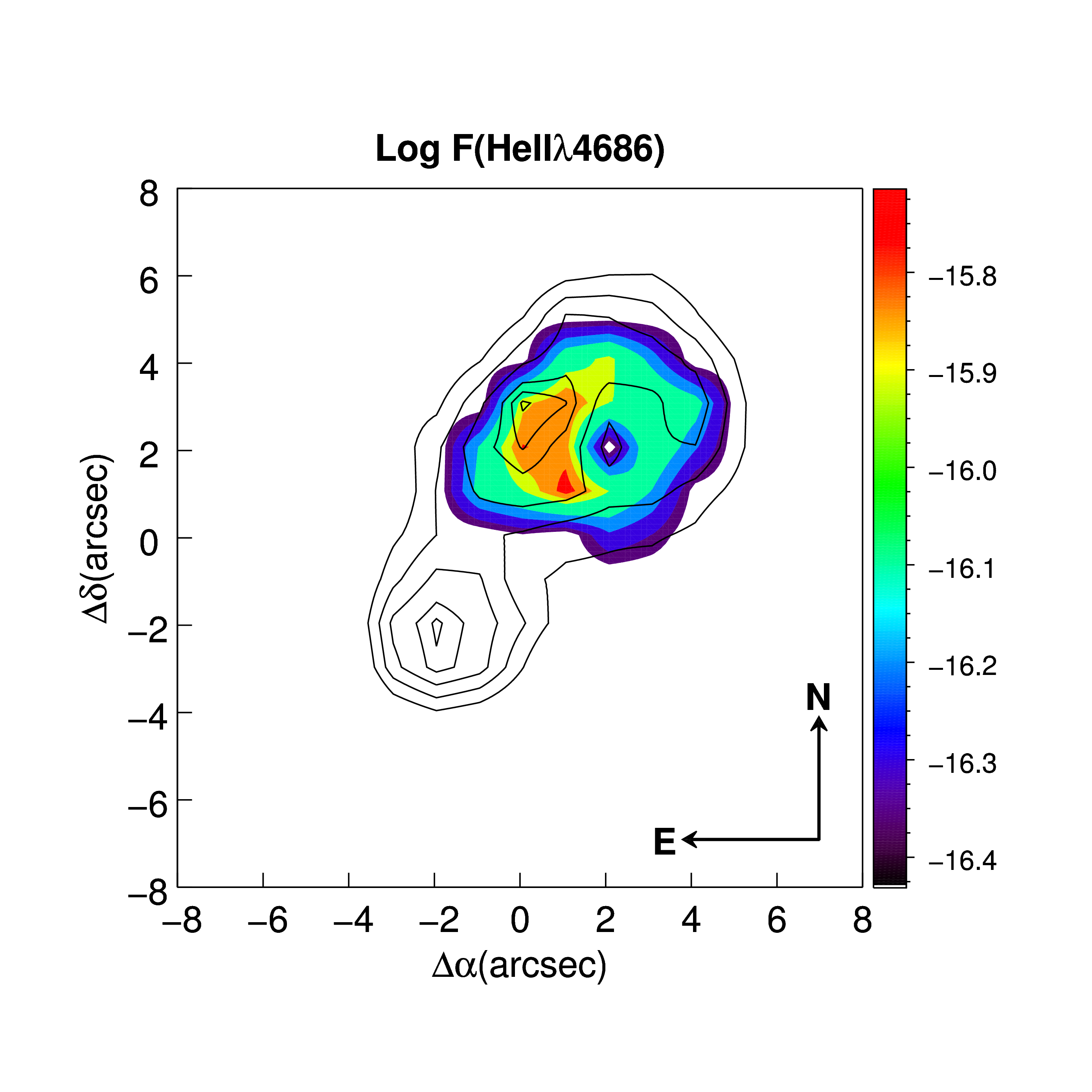}
\includegraphics[width=5cm,clip]{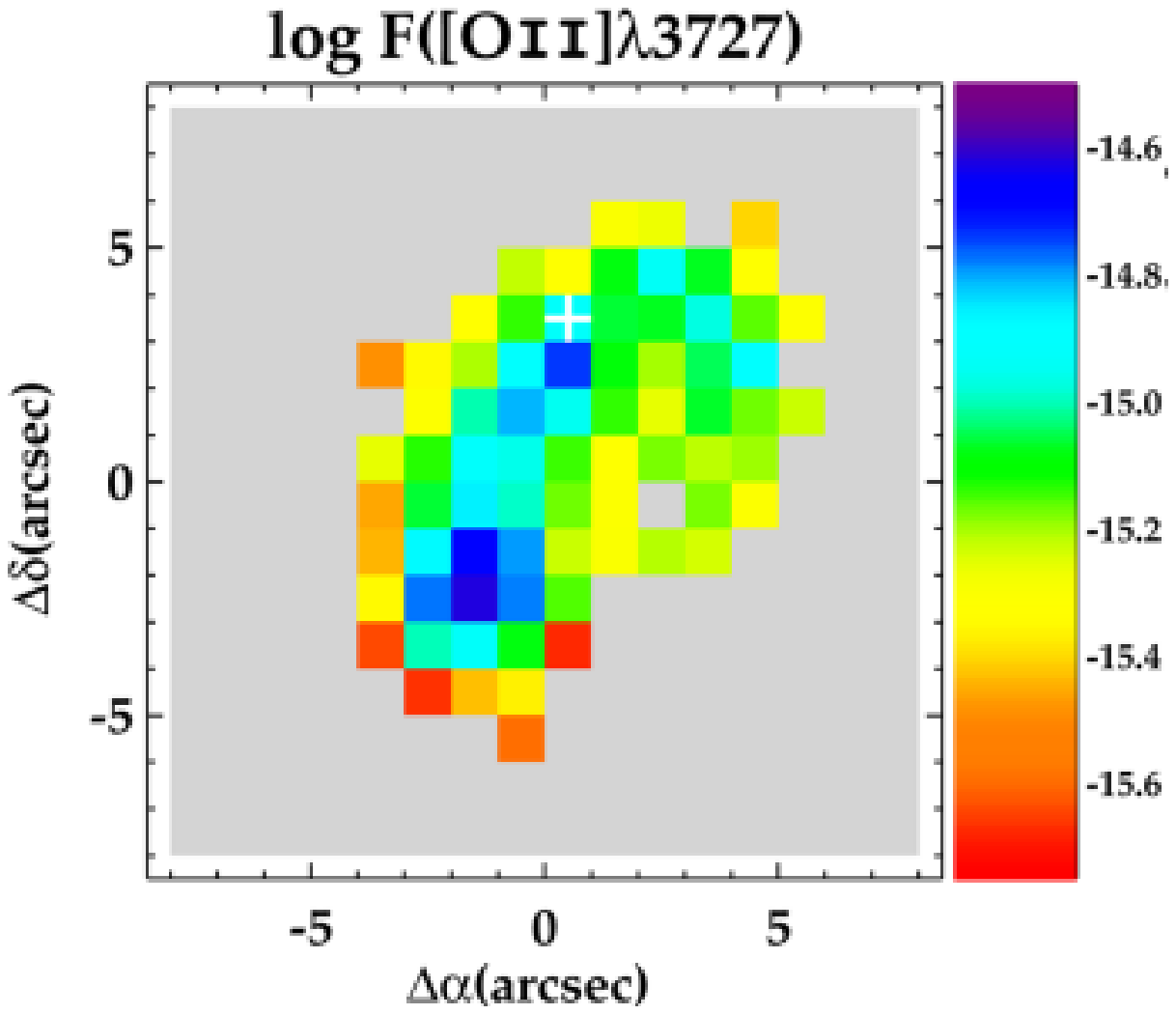}\\  
\includegraphics[width=5cm,clip]{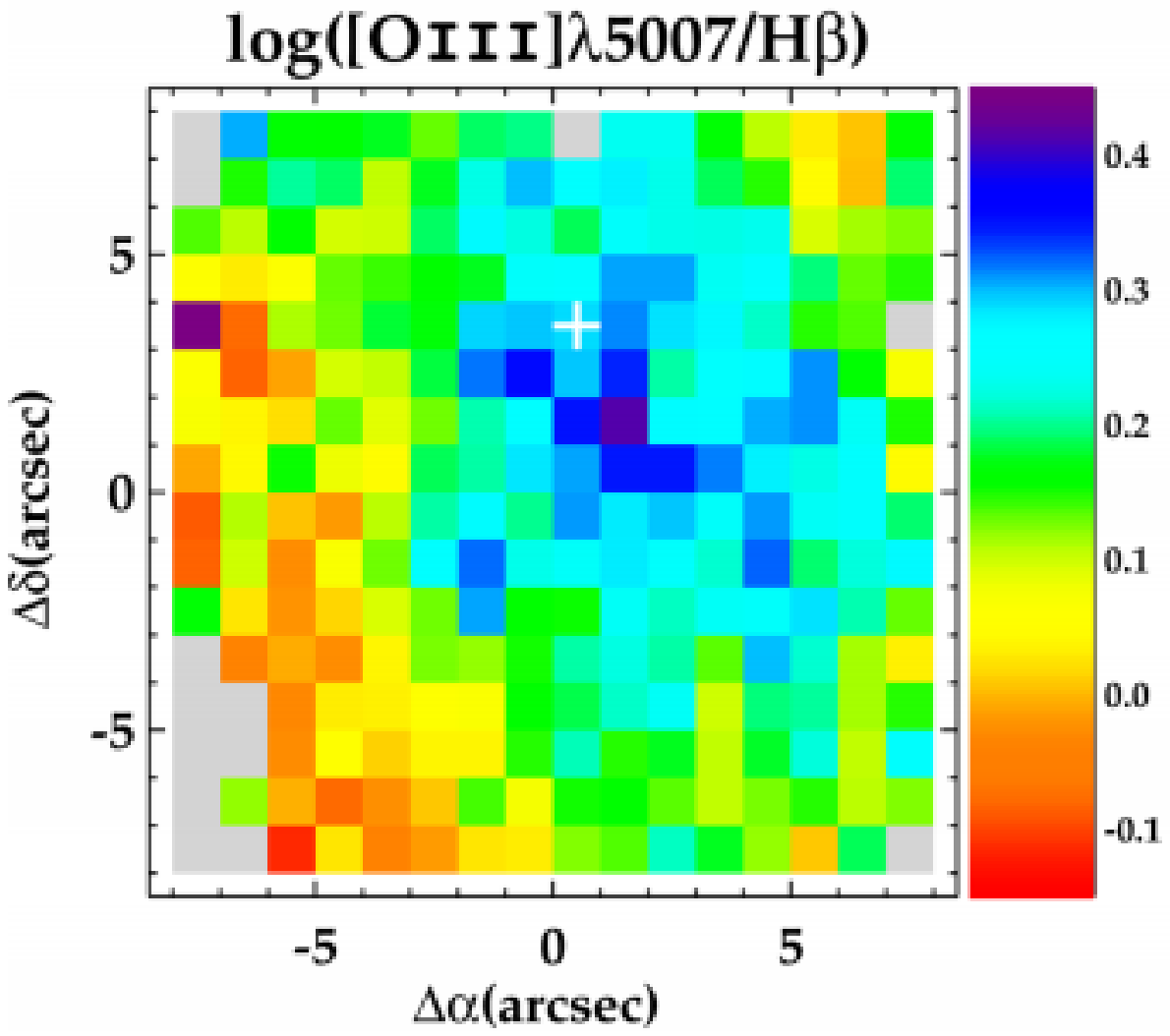}
\includegraphics[width=5cm,clip]{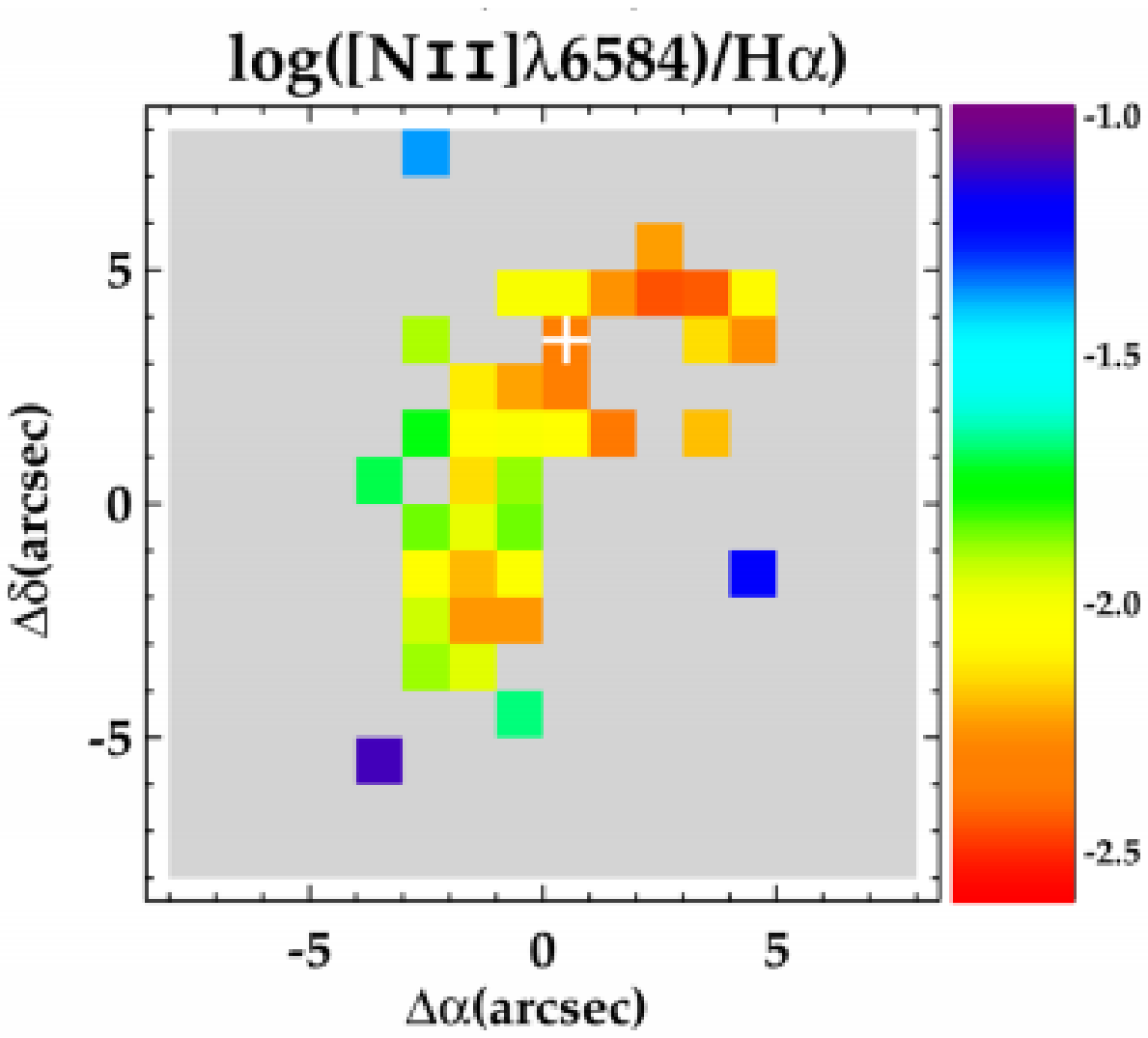}\\
\includegraphics[width=5cm,clip]{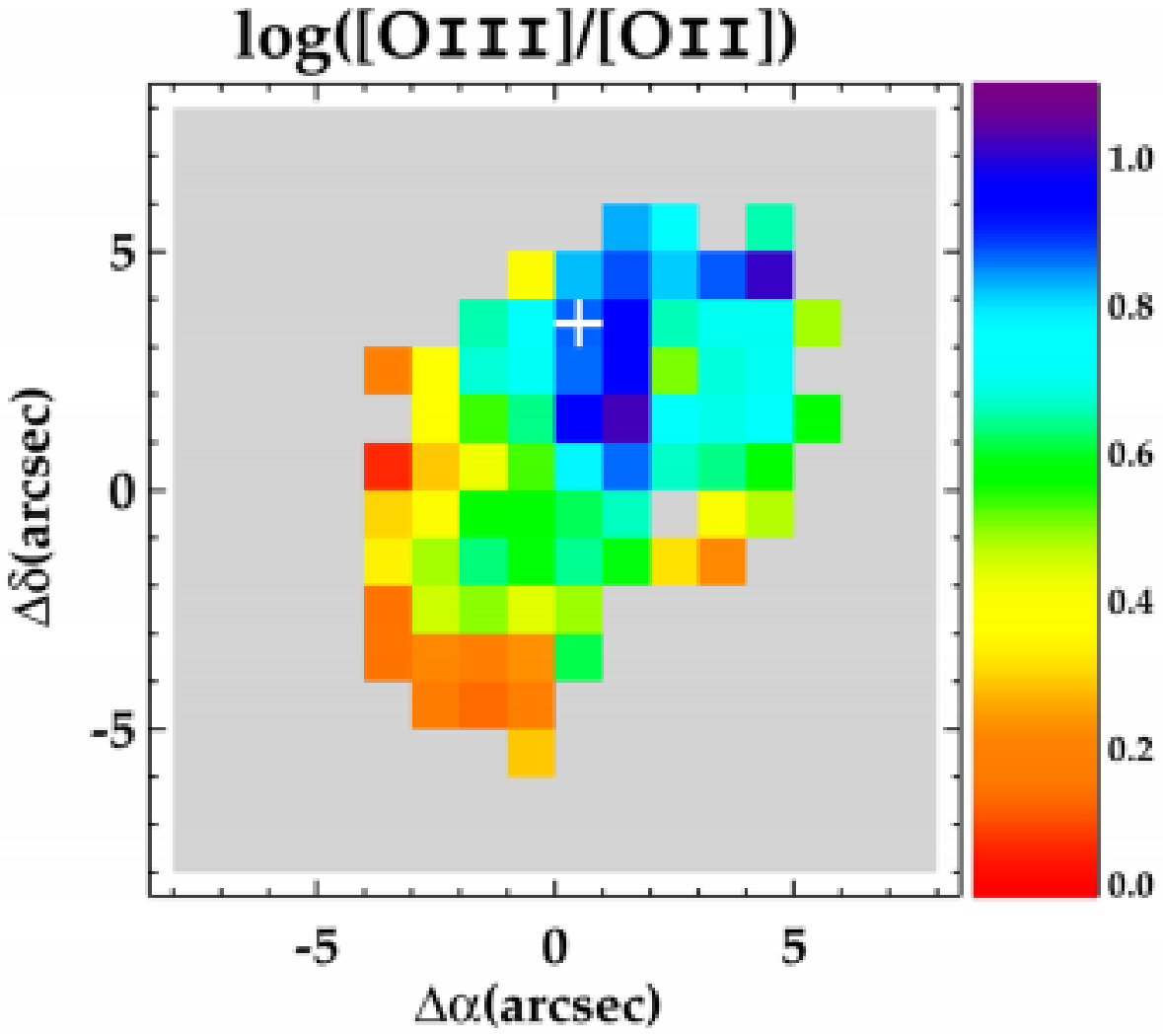}
\includegraphics[width=5cm,clip]{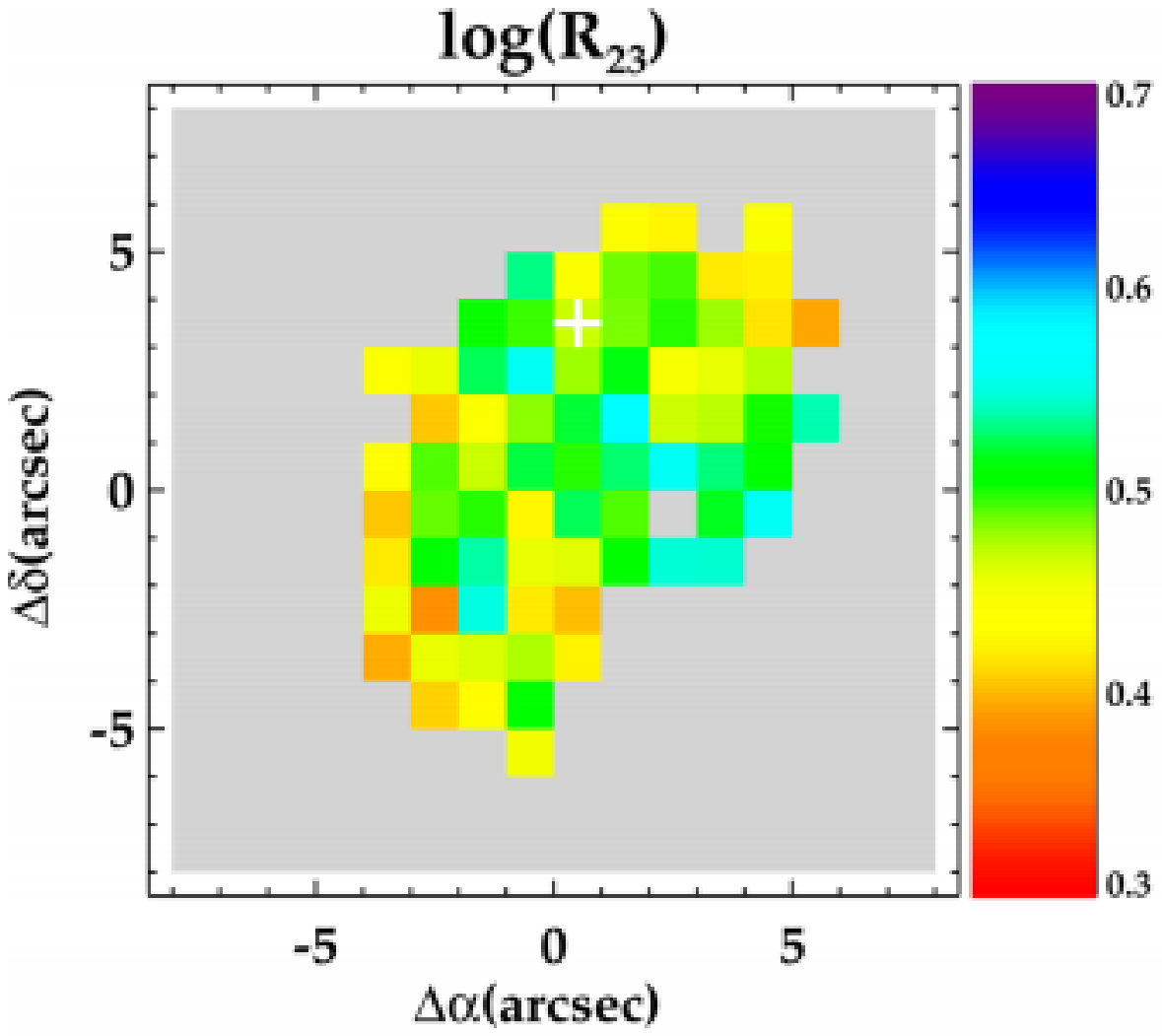}  
\caption{Emssion line flux and line ratio maps: H$\alpha$, [OIII]$\lambda$4363, HeII$\lambda$4686, [OII]$\lambda$3727, [OIII]$\lambda$5007/H$\beta$, [NII]$\lambda$6584/H$\alpha$, [OIII]/[OII], R$_{23}$. Spaxels with no measurements available are left grey. Each spaxel corresponds to 1'' ($\sim$ 88 pc at the distance of 18.2 Mpc). The peak of H$\alpha$  emission is marked with a plus (+) sign on all maps. The cross on the H$\alpha$
map marks the spaxel where we detect the WR feature (see K16). The H$\alpha$ map also shows the boundaries of the areas that we use to create the 1D
spectra of the NW and SE knots, and of the ``plume'' region (see K16 for details). The HeII map is presented as color-filled contour plot and isocontours of the H$\alpha$ emission line are shown overplotted for reference (see K15).  
\label{izw18.fig1}}
\end{center}
\end{figure*}

We performed new IFS observations of IZw18 using the PMAS IFU on the
3.5m telescope at CAHA (see K15, K16). IZw18 is a high-ionization
galaxy which is among the most metal-poor \mbox({Z$\sim$ 1/40
  Z$_{\odot}$}; e.g. Pagel et al. 1992; V{\'{\i}}lchez \& Iglesias-P\'aramo
1998) starbursts in the local Universe. This makes IZw18 an excellent
analog for primeval systems.  Our IFU aperture samples the entire
IZw18 main body and an extended region of its ionized gas (see Fig.\ref{fov}); the two
main SF regions of IZw18 are usually referred to as the north-west
(NW) and southeast (SE) components. We have created and analysed maps
for relevant emission lines, line ratios and physical-chemical
properties of the ionized gas.

\begin{figure*}
\begin{center}
\includegraphics[bb=20 266 565 590,width=0.45\textwidth,clip]{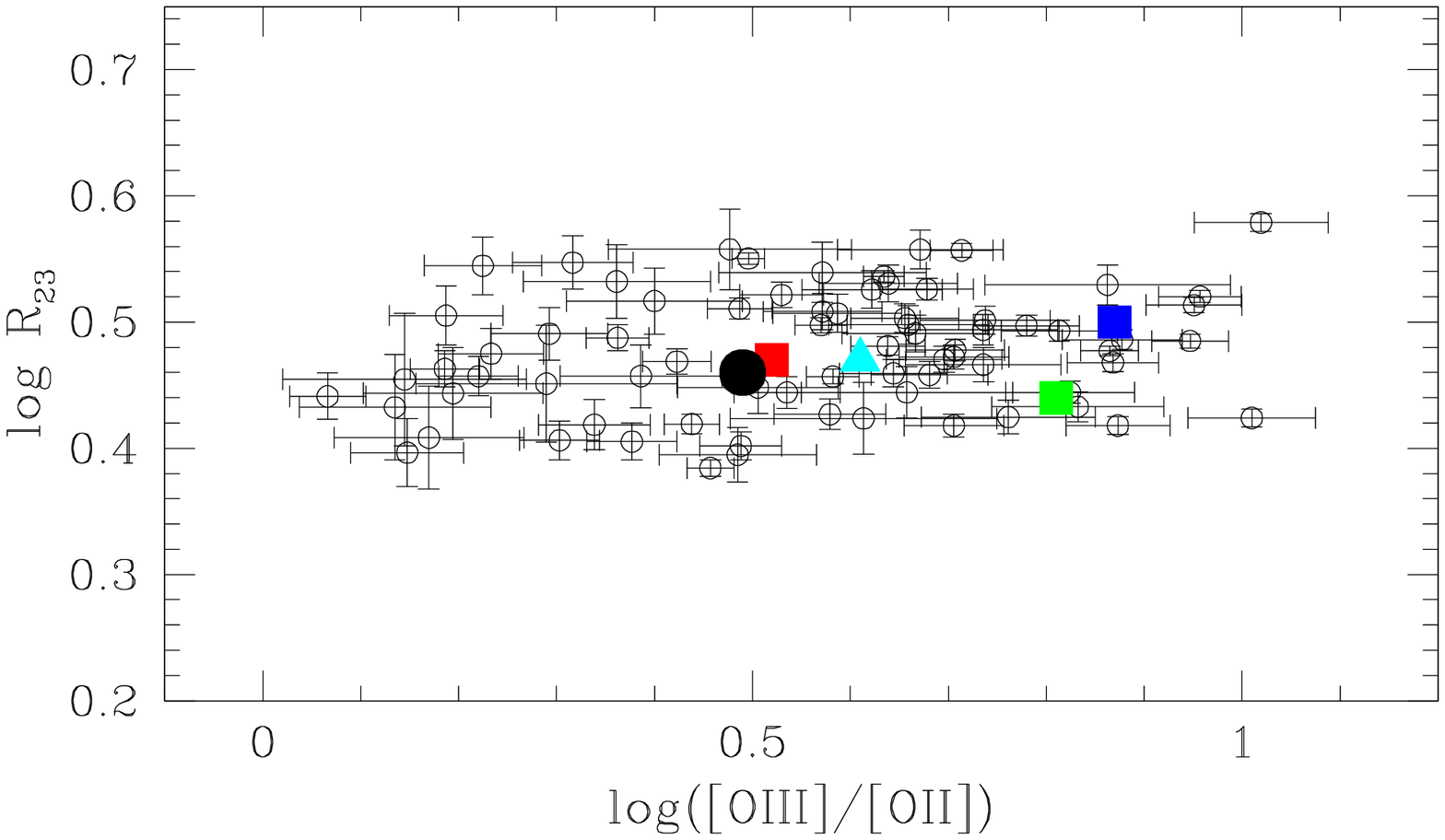}
\includegraphics[bb=20 266 565 590,width=0.45\textwidth,clip]{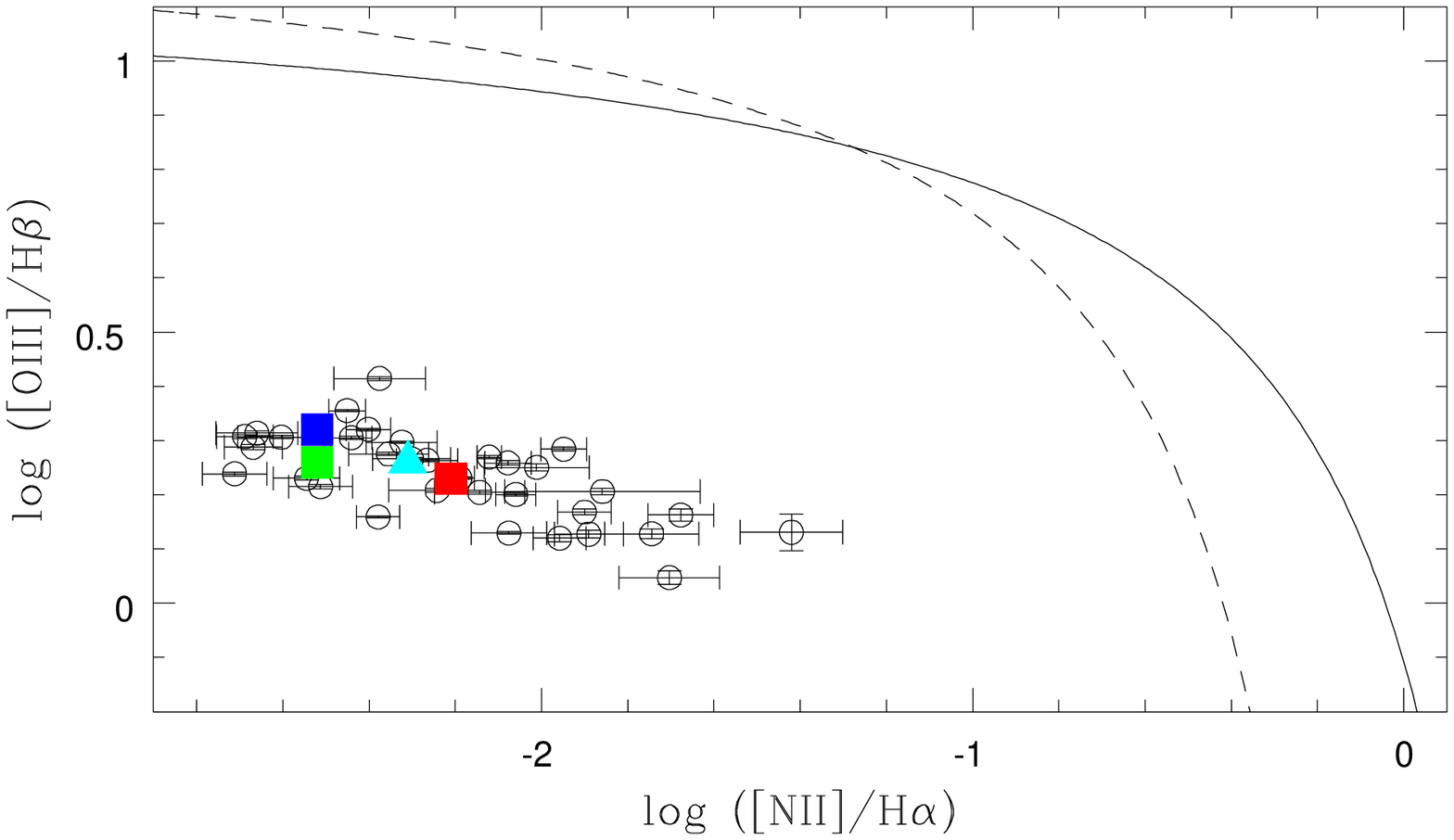}
\caption{{\em Left panel}: The relation between log(R$_{23}$) and log([OIII]/[OII]). Open circles
correspond to individual spaxels from the data cube. {\em Right panel}: The ``BPT'' diagnostic diagram, showing [OIII]$\lambda$5007/H$\beta$ vs. [NII]$\lambda$6584/H$\alpha$. The [SII]/H$\alpha$-BPT and [OI]/H$\alpha$-BPT are shown in K16. Blue, red, and green squares are the line ratios measured from the 1D spectra of the NW knot, SE knot and ``plume'', respectively; the cyan triangle shows
the line ratio values from the total integrated spectrum of IZw18; the black
circle corresponds to the line ratios from the spectrum of the ``halo'' of IZw18 (see K16 for details)
\label{izw18.fig2}}
\end{center}
\end{figure*}

Fig.\ref{izw18.fig1} reveals that the spatial distribution of the
emission in H$\alpha$, HeII$\lambda$4686 and [OIII]$\lambda$4363 is
peaked towards the NW component while the [OII]$\lambda$3727 emission
reaches its maximum at the SE component. Also, by inspecting the maps
of [OIII]$\lambda$5007/H$\beta$, [NII]$\lambda$6584/H$\alpha$ and
[OIII]/[OII], there is a clear tendency for the gas excitation to be
higher at the location of the NW knot and thereabouts. The spatial
distribution of the abundance indicator R$_{23}$ is found to be
relatively flat without any significant peak (see the R$_{23}$ map in
Fig.\ref{izw18.fig1}). However, the ionization parameter diagnostic
[OIII]/[OII] does not show a homogeneous distribution with the highest
values of [OIII]/[OII] found within the NW knot, as mentioned
above. Our spaxel-by-spaxel analysis shows that there is no dependence
between R$_{23}$ and the ionization parameter across IZw18 (see right
panel in Fig.\ref{izw18.fig2}). Other examples of SF regions whith large
range of excitation and constant metallicity can be found in the literature
(e.g. P\'erez-Montero et al. 2011; K13). 

\begin{figure*}
\begin{center}
\includegraphics[bb=1 1 505 405,width=0.45\textwidth,clip]{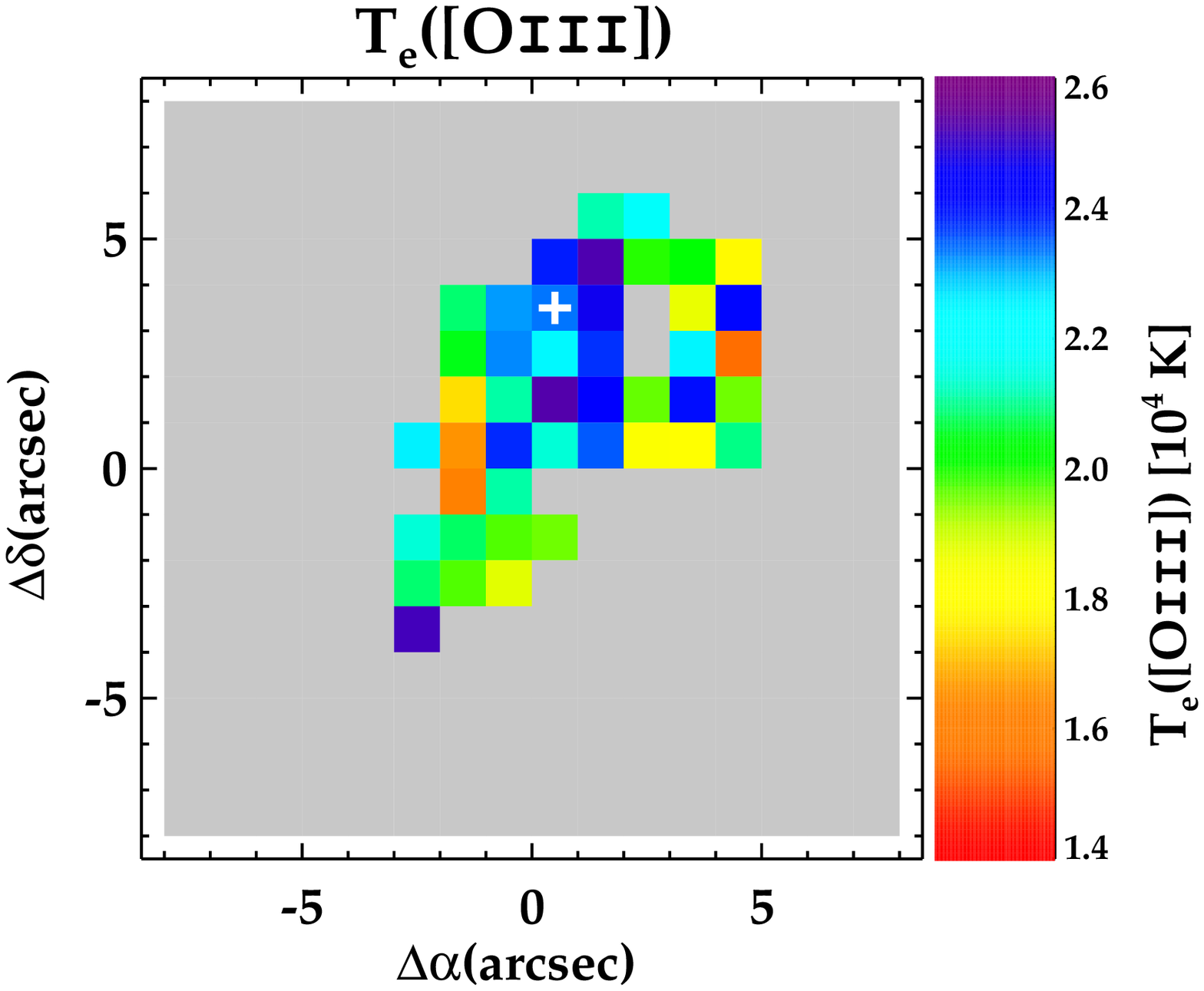}
\includegraphics[bb=1 1 505 405,width=0.45\textwidth,clip]{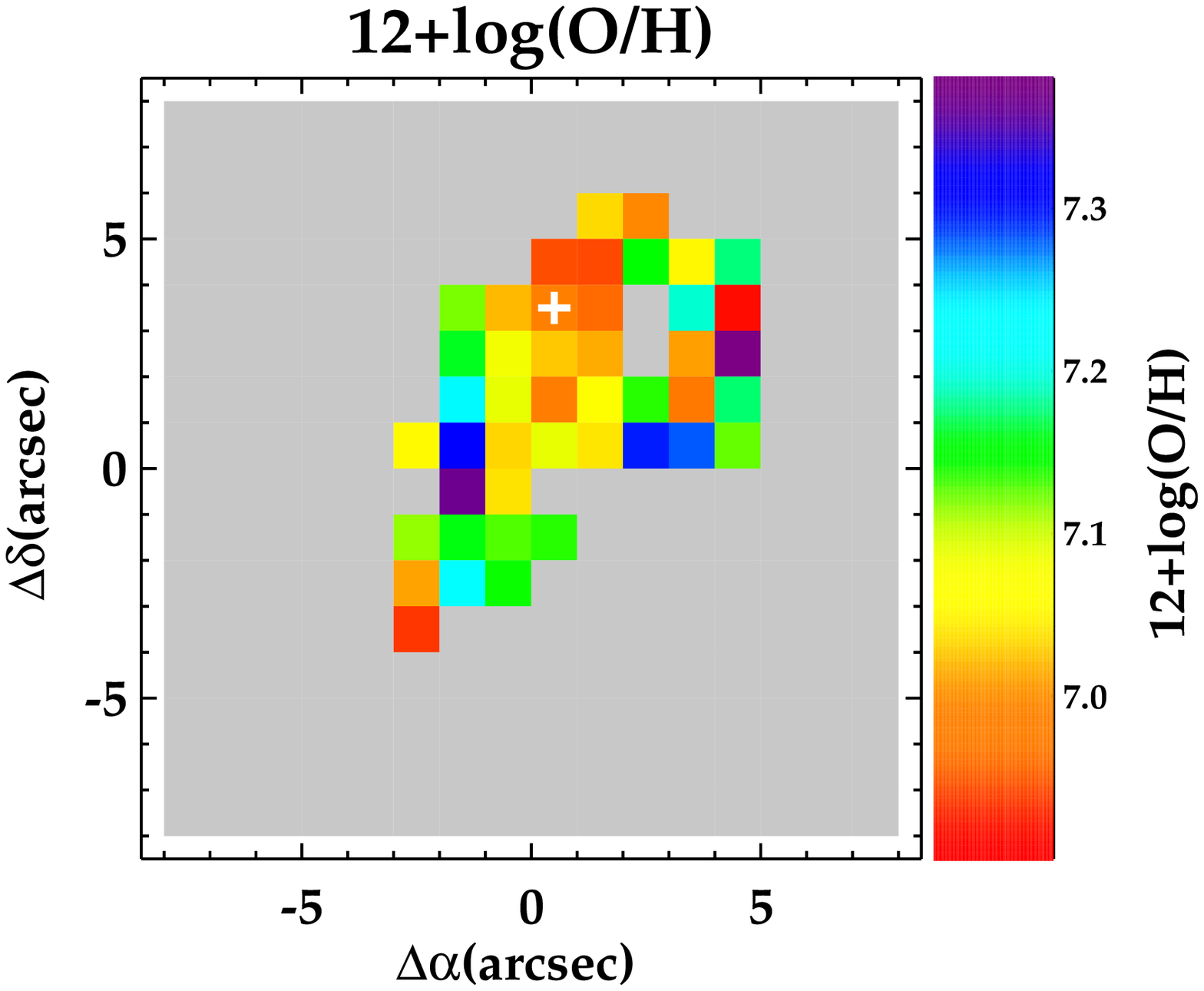} 
\caption{{\em Left panel}: Map of T$_{e}$[OIII] derived directly from the measurement of the [OIII]$\lambda$4363 flux above 3$\sigma$ detection limit. {\em Right panel}: Map of O/H derived from T$_{e}$[OIII].}
\label{izw18.fig3}
\end{center}
\end{figure*}

\begin{figure}
\begin{center}
\includegraphics[bb=20 252 570 590,width=7.5cm,clip]{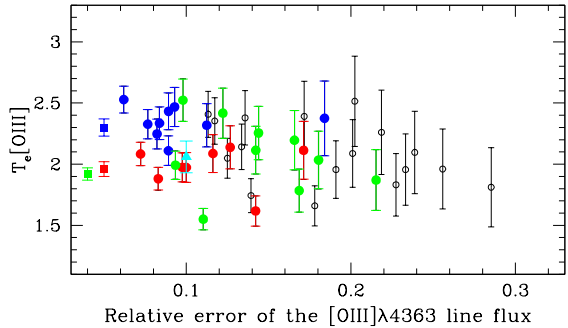} 
\caption{T$_{e}$[OIII] derived directly from the [OIII]$\lambda$4363
  line vs. the relative error of the measurement. Open circles represent individual spaxels; blue, red and green circles indicate the individual spaxels used to create the 1D spectra of the NW knot,
  SE knot, and plume, respectively. Squares indicate the values
  measured from the 1D integrated spectra with the same colour-code as
  used for the individual spaxels.}
\label{izw18.fig4}
\end{center}
\end{figure}

Over $\sim$ 0.30 kpc$^{2}$, using the [OIII]$\lambda$4363 line flux,
we compute T$_{e}$[OIII] values between $\sim$ 15,000-25,000 K (see
Figs.\ref{izw18.fig3} \&~\ref{izw18.fig4}); it is the first time that T$_{e}$[OIII] $>$
22,000 K are derived for IZw18. Fig.\ref{izw18.fig4}
shows that the highest values of T$_{e}$[OIII] are not
an effect of an overestimation during the measurement of the
[OIII]$\lambda$4363 flux. If we look at the ``BPT'' diagram
(Fig.\ref{izw18.fig2}), we
see that all spaxels occupy the SF region, indicating that shocks do
not play an important role in the gas excitation in IZw18. Thus, the enhanced
T$_{e}$[OIII] values derived are expected to be associated primarily with
photoionization from hot massive stars, and T$_{e}$[OIII] errors due
to shocks should be negligible. We note that more than 70$\%$ of the
higher-T$_{e}$[OIII] ($>$ 22,000 K) spaxels are
HeII$\lambda$4686-emitting spaxels too. This reinforces the existence
of a harder ionizing radiation field towards the NW SF knot.

Our statistical analysis shows an important degree of nonhomogeneity
for the T$_{e}$[OIII] distribution and that the scatter in
T$_{e}$[OIII] can be larger than that in O/H within the observed
[OIII] $\lambda$4363-emitting region. We find no statistically
significant variations in O/H (derived directly from T$_{e}$[OIII]) across the PMAS-IFU aperture, indicating a global
homogeneity of the O/H in IZw18 over spatial scales of hundreds of
parsecs. The representative metallicity of IZw18 derived here, from
individual spaxel measurements, is 12 + log(O/H) = 7.11 $\pm$ 0.01
(error-weighted mean value of O/H and its corresponding statistical
error). The prevalence of a substantial degree of homogeneity in O/H
over IZw18 can constrain its chemical history, suggesting
an overall enrichment phase previous to the current burst.

We took advantage of our IFU data to create 1D integrated spectra for
regions of interest in the galaxy. For the first time, we derive the
IZw18 integrated spectrum by summing the spaxels over the whole
FOV. Physical-chemical properties of the ionized gas were derived from
these selected region spectra.  We also show that the derivation of
O/H and N/O does not depend on the aperture size used. This is a
relevant result for studies of high-redshift SF objects for which
only the integrated spectra are available.

{\bf PopIII-star siblings: a possible culprit behind the extended
  nebular HeII$\lambda$4686 emission in IZw18} 

Narrow HeII emission in SF
galaxies has been suggested to be mainly associated with
photoionization from WRs, but WRs cannot satisfactorily explain the
HeII-ionization in all cases, particularly at lowest metallicities
where nebular HeII emission is often seen and observed to be stronger
(e.g. Kehrig et al. 2004; Shirazi \& Brinchmann 2012; James et
al. 2016). 
Why is studying the formation of HeII emission relevant ?
HeII emission indicates the presence of high energy photons (E $\geq$
54 eV), and so provides essential constraints on the SEDs of hot
massive stars.  HeII-emitters are apparently more frequent among
high-z galaxies than for local objects (Kehrig et al. 2011;
Cassata et al. 2013). Actually, narrow HeII emission has been claimed to be a
good tracer of PopIII-stars (the first very hot metal-free stars) in
distant galaxies (e.g. Schaerer 2003); these stars are believed to
have contributed significantly to the reionization of the Universe, a
challenging subject in contemporary cosmology. In fact, searching for
PopIII-hosting galaxy candidates is
among the main science drivers for next generation telescopes (e.g.,
JWST; E-ELT). However, we should note that the origin of narrow HeII lines remains difficult to
understand in many nearby and distant SF galaxies/regions (e.g.,
Kehrig et al. 2011; Shirazi \& Brinchmann 2012; Gr\"afener \& Vink 2015; Pallottini et al. 2015; Hartwig et
al. 2016). One of the main reasons why we do not fully understand the
physics behind the formation of nebular HeII is the lack of direct
probes on HeII-ionizing hot stars. Detailed investigation of the HeII
emission at low redshift is needed to better interpret distant
narrow HeII-emitters and therefore gaining a deeper understanding of the reionization epoch. IZw18, as the most metal-poor HeII-emitter
in the local Universe, is an ideal object to perform this study.

Our IFS data reveal for the first time the entire nebular
HeII$\lambda$4686-emitting region (see map of HeII$\lambda$4686 in
Fig.\ref{izw18.fig1}) and corresponding total HeII-ionizing photon
flux \mbox{[Q(HeII)$_{obs}$]} in IZw18.  These observations combined
with stellar model predictions point out that conventional excitation
sources (e.g., single WRs, shocks, X-ray binaries) cannot convincingly
explain the total Q(HeII)$_{obs}$ derived for IZw18 (see K15
for details).  Other mechanisms are probably also at work. If the
HeII-ionization in IZw18 is due to stellar sources, these might be
peculiar very hot stars. Based on models of very massive O stars
(Kudritzki 2002), $\sim$ 10-20 stars with 300 M$_{\odot}$ at
Z$_{IZw18}$ [or lower, down to Z$\sim$(1/100) Z$_{\odot}$] can
reproduce our total Q(HeII)$_{obs}$ (see also Sz\'ecsi et
al. 2015). However, the super-massive star scenario requires a cluster
mass much higher than the mass of the IZw18 NW knot (where the HeII
region is located), and it would not be hard enough to explain the
highest HeII/H$\beta$ values observed. On top of that, the existence
of super-massive 300 M$_{\odot}$ stars remains heavily debated, and an
extrapolation of the IMF predicting such massive stars remains
unchecked up to now (Vink et al. 2014).

Considering that the previous scenarios fail in reproducing the
observations, and that searches for PopIII-hosting galaxies have been
carried out using HeII lines (e.g. Schaerer 2008; Cassata et
al. 2013), we thought that (nearly) metal-free hot stars may hold the
key to the HeII ionization in IZw18. To test this hypothesis, as an
approximation of (nearly) metal-free stars in IZw18 -- the so-called
{\it PopIII-star siblings} -- we compared our observations with
current models for rotating Z=0 stars (Yoon et al. 2012), which in
fact reproduce our data better: $\sim$8-10 of such stars with
M$_{ini}$=150 M$_{\odot}$ can explain the total Q(HeII)$_{obs}$ and
the highest HeII/H$\beta$ values observed.  The PopIII-star sibling
scenario, invoked for the first time in IZw18 by K15, goes in line
with the results by Lebouteiller et al. (2013). While gas in IZw 18 is
very metal-deficient but not primordial, Lebouteiller et al. (2013)
have pointed out that the HI envelope of IZw 18 near the NW knot
contains essentially metal-free gas pockets. These gas pockets could
provide the raw material for making such {\it PopIII-star siblings} (see also Tornatore et al. 2007).

\section{Summary and Conclusions}
\label{discussion}

A brief overview of the first optical IFU observations of two nearby,
extremely metal-poor {\em bursty}-galaxies (Mrk178 and IZw18) is
presented.  Clues of the early-universe can be found in our cosmic
backyard through this class of objects which are excellent primordial
analogues, and are key in understanding galaxy evolution.  IFS studies
of such galaxies enable extended insight into their ``realistic'' ISM
and massive stars, therefore providing constraints on high-redshift
galaxy evolution, and on metal-poor stellar models. Our data provide
useful testbench for realistic photoionization models at the lowest
metallicity regime.

The elusive PopIII-hosting galaxies have been searched through the
high-ionization HeII line signature. The HeII line is in comfortable
reach of next generation telescopes, like JWST and TMT, which will
detect the rest-frame UV of thousands of galaxies during the epoch of
reionization. In light of these new observations, a more sophisticated
understanding of the high-ionization phenomenon is required to
interpret the data in a physically meaningful manner, and to possibly
constrain sources responsible for the Universe reionization. Using IFU
data, we were able to recover the {\em total} HeII luminosity and
perform a {\em free-aperture} investigation on the formation of narrow
HeII line. Our observations of IZw18 test the current
poorly-constrained models for metal-poor massive stars, and suggest
that peculiar hot stellar sources, as PopIII-star siblings, might be
culprits for the HeII-ionization in this galaxy. This result
emphasizes the need to identify extremely metal-poor
HeII-emitting targets, such as IZw18, and the power of IFS for such
kind of investigation.

\acknowledgments CK gratefully acknowledges the co-authors and referees from K13,
K15 and K16 who made significant contribution to the results presented
in this review. 

\end{document}